\documentclass[journal=jctcce,manuscript=article]{achemso}

\usepackage[version=3]{mhchem} % Formula subscripts using \ce{}
\usepackage[english]{babel}
\usepackage{graphicx}% Include figure files
\usepackage{caption}
\usepackage{subcaption}
\usepackage[mathlines]{lineno} % Enable numbering of text and display math
\usepackage{amsmath}
\usepackage{amsfonts}
\usepackage{amssymb}
\author{Jayashrita Debnath}
 \affiliation{Department of Theoretical Biophysics, Max Planck Institute of Biophysics, 60438 Frankfurt am Main, Germany }
\author{Gerhard Hummer}
\email{gerhard.hummer@biophys.mpg.de }
\affiliation{Department of Theoretical Biophysics, Max Planck Institute of Biophysics, 60438 Frankfurt am Main, Germany }
\altaffiliation{Institute of Biophysics, Goethe University Frankfurt, 60438 Frankfurt am Main, Germany}

\title{Random functions as data compressors for machine learning of molecular processes} %Title of paper

\begin{document}

\begin{abstract}
 Machine learning (ML) is rapidly transforming the way molecular dynamics simulations are performed and analyzed, from materials modeling to studies of protein folding and function. ML algorithms are often employed to learn low-dimensional representations of conformational landscapes and to cluster trajectories into relevant metastable states. Most of these algorithms require selecting a small number of features that describe the problem of interest. Although deep neural networks can tackle large numbers of input features, the training costs increase with input size, which makes the selection of a subset of features mandatory for most problems of practical interest. Here, we show that random nonlinear projections can be used to compress large feature spaces and make computations faster without substantial loss of information. We describe an efficient way to produce random projections and then exemplify the general procedure for protein folding. For our test cases NTL9 and the double-norleucin variant of the villin headpiece, we find that random compression retains the core static and dynamic information of the original high dimensional feature space and makes trajectory analysis more robust.
\end{abstract}

\section{Introduction}
Molecular dynamics (MD) simulations have proven to be a very useful tool for the study of biomolecular systems. Large systems with millions of atoms are now frequently simulated for microseconds to milliseconds \cite{Wilson2021,Sikora2021,Casalino2020}. This remarkable progress has, however, led to a new challenge: the problem of analyzing long trajectories of high dimensional data \cite{Jung2023}.

Although the dimensionality of each frame of an MD trajectory scales with the number of particles in the box, this dimensionality of the trajectories can be reduced due to the inherent timescale separation of the encoded dynamics\cite{Best2005,Krivov2008,Bittracher2017}. For instance, ions and water relax much faster than protein conformations. The dynamics of the protein, too, can be separated into the slower global movements of domains and faster fluctuations in the flexible regions. These observations have motivated researchers to eliminate degrees of freedom considered fast and non-essential. Such elimination typically begins with ignoring solvent degrees of freedom. Following this step, it has been a common practice to project the dynamics of the trajectory onto a few collective variables (or order parameters) such as the root mean squared deviation(RMSD) from a structure, dihedral angles, or distances of interest. However, with increasing size and complexity of the systems, simplistic elimination strategies do not work well any more, not least because the slow degrees of freedom become less obvious. Consequently, machine learning techniques are being routinely employed to reduce the dimensionality of MD trajectories.

Machine learning based approaches not only help in learning meaningful lower dimensional representations of trajectories, they also help in clustering trajectories into metastable states or learning collective variables for enhanced sampling methods\cite{Tribello2019,Sun2022,No2020,Jung2023,encodermap2,glielmo_review_md_data}. In practical applications, one usually starts by choosing a set of input features for training the model. In systems like proteins in a box of water, the water molecules are often neglected and trajectories are represented using internal coordinates of the protein atoms such as C$\alpha$ distances (or contacts), dihedral angles of the residues or cartesian coordinates of a subset of atoms. The goal of any dimensionality reduction technique then is to find an $n$ dimensional map $\Phi$ of the original $N$ dimensional feature space, $ \Phi: \mathbb{R}^N \mapsto \mathbb{R}^n$ where $n\ll N$, which resolves relevant states and is associated with a simple, near-Markovian dynamics. Linear maps are represented by a $n\times N$ matrix $\mathcal{M}$. Many strategies have been employed over the years for generating such mappings\cite{Nedialkova2014}. One of the most popular algorithms used for MD trajectory data is Principal Component Analysis (PCA)\cite{garcia-1992, Palma2022}, in which a linear combination of the initial feature space is obtained in a way that optimally describes the variance of the data. Another dimensionality reduction technique that is often employed in the context of time dependent data is Time-structure Independent Component Analysis (TICA) where the data are projected onto the generalized eigenvectors of time-lagged covariance matrix, accounting for static correlations, to separate slow from fast relaxation processes\cite{PrezHernndez2013} that generated it. While PCA, TICA or other approaches like Linear Discriminant Analysis (LDA) often result in meaningful lower dimensional representations of the data, these linear methods often fail to provide meaningful lower dimensional representation of the data when the dimensionality of input feature space is very high\cite{Doerr2017}.

For (bio)molecular systems, agnostic feature spaces are large. The number of pair distances scales quadratically with the number of residues in a protein, and the number of dihedral angles scales linearly. Such features thus cannot be used directly as input for most linear machine learning algorithms and a reduction of dimensions becomes necessary even before the application of these techniques. Some nonlinear dimensionality reduction techniques such as t-distributed Stochastic Neighbour Embedding (t-SNE)\cite{Spiwok2020}, Kernel PCAs, Self organizing maps\cite{Bouvier2014}, Isomaps\cite{Tenenbaum2000-mv}, Sketch map\cite{Tribello2012}, Encodermap\cite{lemke_encodermap_2019}, VAMPnet\cite{VAMPNet} are increasingly being employed for analyzing MD simulations as they can deal with input of higher dimension\cite{Tribello2019}. However, these methods too cannot deal with excessively large input dimensionality, as would be the case when solvent degrees of motion are included or when proteins are not small. In such cases, discarding some input features is imperative, even when working with nonlinear models for dimensional reduction.

Recently, a lot of work has focused on reducing input feature sizes by extracting a subset of useful features from the larger set\cite{Mosaic,AMINO,Rydzewski2023,Rydzewski2023_slowkin}. These strategies often aim to remove redundant features using some variant of mutual information based metric. In practice, it has been found that although these methods are very good in removing inessential features, they tend to become intractable or computationally expensive for larger data sets.

Here, we propose an alternative approach to achieving reduced input feature sizes. In contrast to the existing methodologies, we compress the large features using random nonlinear projections. Random projections could be an efficient strategy to produce compressed feature sets before the application of any machine learning algorithm for analyzing molecular dynamics trajectories. In the following sections, we introduce random projections, propose a way to generate random nonlinear projections for MD trajectory data, and demonstrate that such compressed feature sets preserve essential properties of the original high dimensional data for protein-folding studies.

\section{Methods}
Our strategy to generate random nonlinear projections is inspired by the Transition Manifolds method\cite{Bittracher2017,Bittracher2020} and the Whitney Embedding Theorem \cite{Whitney1936}, two approaches that together guarantee the existence of a lower dimensional embedding for MD trajectory data. Mathematically, our random nonlinear projection approach can be seen as an extension of the random mappings method, a linear dimensionality reduction technique, that was initially proposed and applied in the context of document classification. In the random mappings method, a lower dimensional map is generated using a $n\times N$ matrix $\mathcal{M}$ that is randomly initialized. A linear random mapping is given by $\boldsymbol{x}_{n \times T} = \mathcal{M}_{n \times N} \boldsymbol{X}_{N \times T}$, where $\boldsymbol{X}$ is a matrix of dimensions $N\times T$ with $N$ the number of input features and $T$ the length of the trajectory data, and $\boldsymbol{x}$ is a $n\times T$ matrix representing the input data mapped into a space of dimension $n$ ($n \leq N$)\cite{Bingham2001}.

If the column vectors of the random matrix $\mathcal{M}_{n \times N}$ are drawn from a mean-free, unit-variance distribution, the resulting random combinations of the original high dimensional features are almost orthogonal\cite{HechtNielsen94ContextVectors,Kaski,dasguptaRandProj}. Additionally, when the lower dimension is sufficiently large these mappings preserve well all the pairwise distances between the original data (Johnson-Lindenstraus lemma\cite{Dasgupta2002}). Reducing the dimensionality using these mappings can therefore speed up classification or clustering tasks while causing almost no loss of information as long as a the embedded dimension $n$ is sufficiently large\cite{Wjcik2018,Bingham2001}. However, the required dimension $n$ is extremely large for typical MD trajectory data. Many other strategies have thus been proposed for generating random projections \cite{Li2007,Achlioptas2001} including approaches to generate random nonlinear projections\cite{theory_random_projections}. In the following paragraphs, we propose our strategy to generate random nonlinear projections for MD trajectory data that falls into the same category as the latter approaches.

\subsection{Constructing random projections of MD data}
To generate compressed feature spaces, we perform a forward propagation of our high dimensional trajectory data through randomly initialized feed forward networks. Generating these projections using a single layer perceptron would be mathematically equivalent to the linear random mapping method due to the absence of any nonlinear activation. However, the final projected space generated using multi-layer perceptrons with nonlinear activations is nonlinear even in the absence of any bias. An $n$-dimensional nonlinear embedding can therefore, be generated either using $n$ different multi-layered networks, each with a single output neuron, or from a single multi-layered network with $n$ output neurons.

\begin{figure}[htbp]
    \centering
    \includegraphics[scale=0.3]{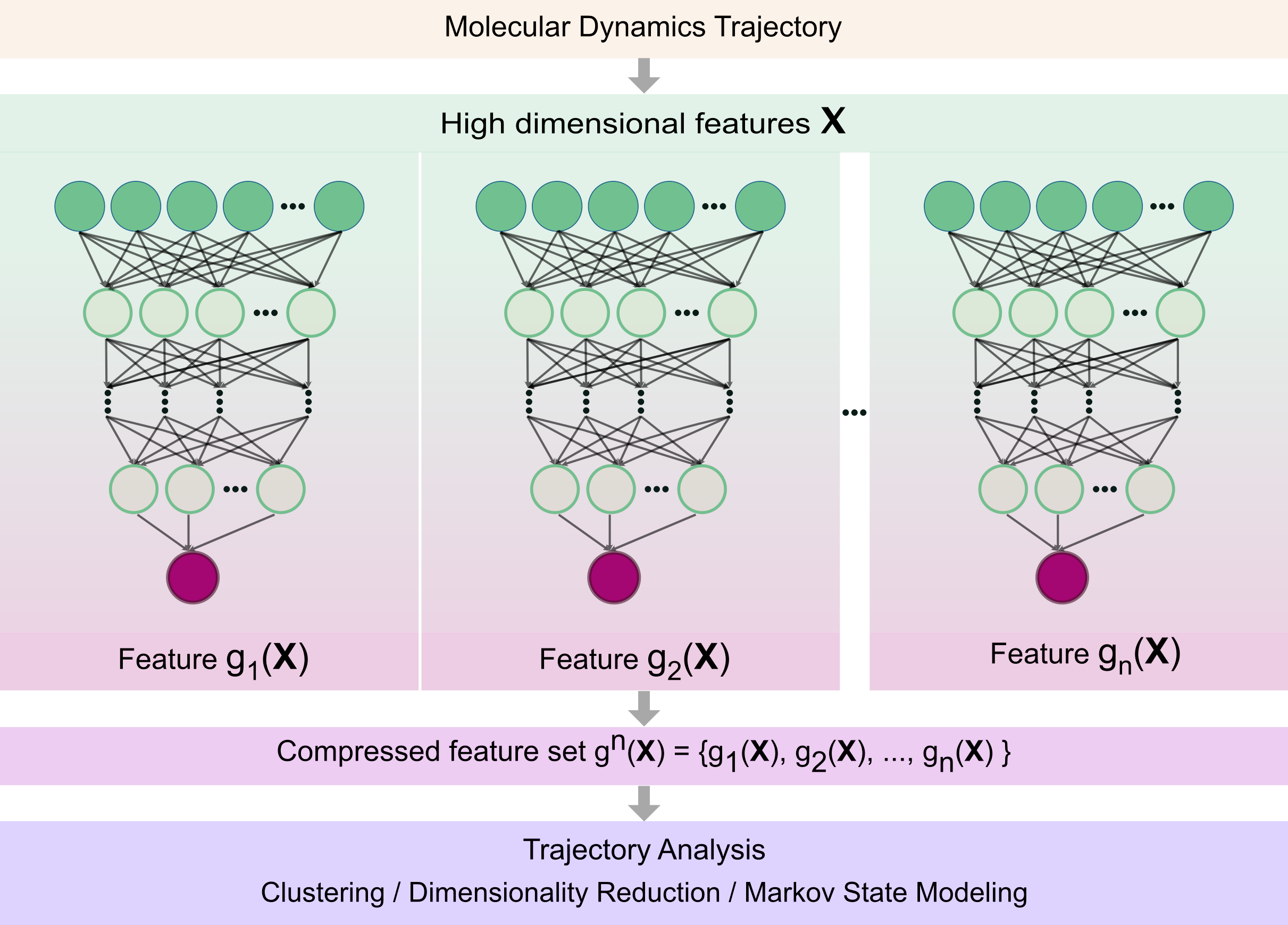}
    \caption{Random compression of MD trajectories. Vectors $\boldsymbol{X}$ containing $N$ molecular dynamics features of the structures along a MD trajectory are compressed to dimension $n\ll N$ using different random networks $g_i(\boldsymbol{X})$ with $i=1,\ldots,n$. The resulting $n$ low-dimensional projections are then used for further trajectory analysis.}
    \label{fig:network}
\end{figure}

Networks with one output have an architecture as shown in Figure \ref{fig:network}, with a random number of hidden layers and a random width for each of the hidden layers. Mathematically, such a transformation is equivalent to:
\begin{align}
g_{\alpha}^{(0)}({\boldsymbol{X}}) &= {\boldsymbol{W}}_{\alpha}^{(0)}{\boldsymbol{X}} + B_{\alpha}^{(0)} \\
g_{\alpha}^{(1)}({\boldsymbol{X}}) &= \phi_{\alpha}^{(1)}\big({\boldsymbol{W}}_{\alpha}^{(1)} g_{\alpha}^{(0)}({\boldsymbol{X}}) + B_{\alpha}^{(1)}\big) \\
g_{\alpha}^{(h_{\alpha}-1)}({\boldsymbol{X}}) &= \phi_{\alpha}^{(1)}\big({\boldsymbol{W}}_{\alpha}^{(h_{\alpha}-1)} g_{\alpha}^{(h_{\alpha}-2)}({\boldsymbol{X}}) + B_{\alpha}^{(h_{\alpha}-1)}\big) \\
g_{\alpha}^{(h_{\alpha})}({\boldsymbol{X}}) &= {\boldsymbol{W}}_{\alpha}^{(h_{\alpha})} g_{\alpha}^{(h_{\alpha}-1)}({\boldsymbol{X}}) + B_{\alpha}^{(h_{\alpha})}
\label{Eq: randomfunctions}
\end{align}
where ${\boldsymbol{W}}_{\alpha}^{(i)}{\boldsymbol{X}}$, $B_{\alpha}^{(i)}$ are weights and biases of the $i^{th}$ layer, $g_{\alpha}^{(h_{\alpha})}({\boldsymbol{X}})$ is a one-dimensional vector obtained using $h_{\alpha}$ hidden layers activated by ELU activations, $\phi_{\alpha}^{(i)}$, after each hidden layer $i$ for a given network $\alpha$. The outputs $g_{\alpha}^{(h_{\alpha})}({\boldsymbol{X}})$ are standardized using min-max normalization. An $n$ dimensional random nonlinear projection, $g^{n}({\boldsymbol{X}})$, is then obtained by generating $n$ random function vectors $\{ g_{1}^{(h_{1})}({\boldsymbol{X}}), g_{2}^{(h_{2})}({\boldsymbol{X}}), \dddot{}, g_{n}^{(h_{n})}({\boldsymbol{X}}) \}$. As these networks are not trained, the method used for initializing the weights and biases influences the quality and stability of projections obtained. In following sections, we have used Xavier initialization scheme for initializing the weights of the networks while the values of the biases were initialized from a uniform distribution.

A good compressed feature set should ideally include a diverse set of features that are not highly correlated. In practice, using a single network often results in many correlated functions as output. However, when multiple networks having the same or different architectures (varying the number and depth of the hidden layers) are used to generate different one dimensional embeddings, the resulting random functions are often less correlated. In the following sections, we restrict our discussion to compressed feature spaces that are generated using multiple independent networks.

As discussed earlier, many artificial neural network based methods have been developed recently to learn reaction coordinates or committors from MD trajectories\cite{Jung2023,Belkacemi2021,Chen2018,Bonati2021,Chen2019,Hernndez2018,Naleem2023}. By contrast, here we only intend to compress the high dimensional space to make any further analysis more tractable. The best lower dimensional representation or reaction coordinate might not be obtained in this process. However, the compressed space, having higher dimensionality than the best lower dimensional representation, should still be able to retain all relevant kinetic and metastable state information in order to be effective. Having proposed a way to generate compressed feature spaces, we now assess their ability to retain timescales and clusters for different systems in the following sections.

\section{Results}
\subsection{Alanine dipeptide}
\begin{figure}[htbp]
    \centering
    \includegraphics[scale=0.5]{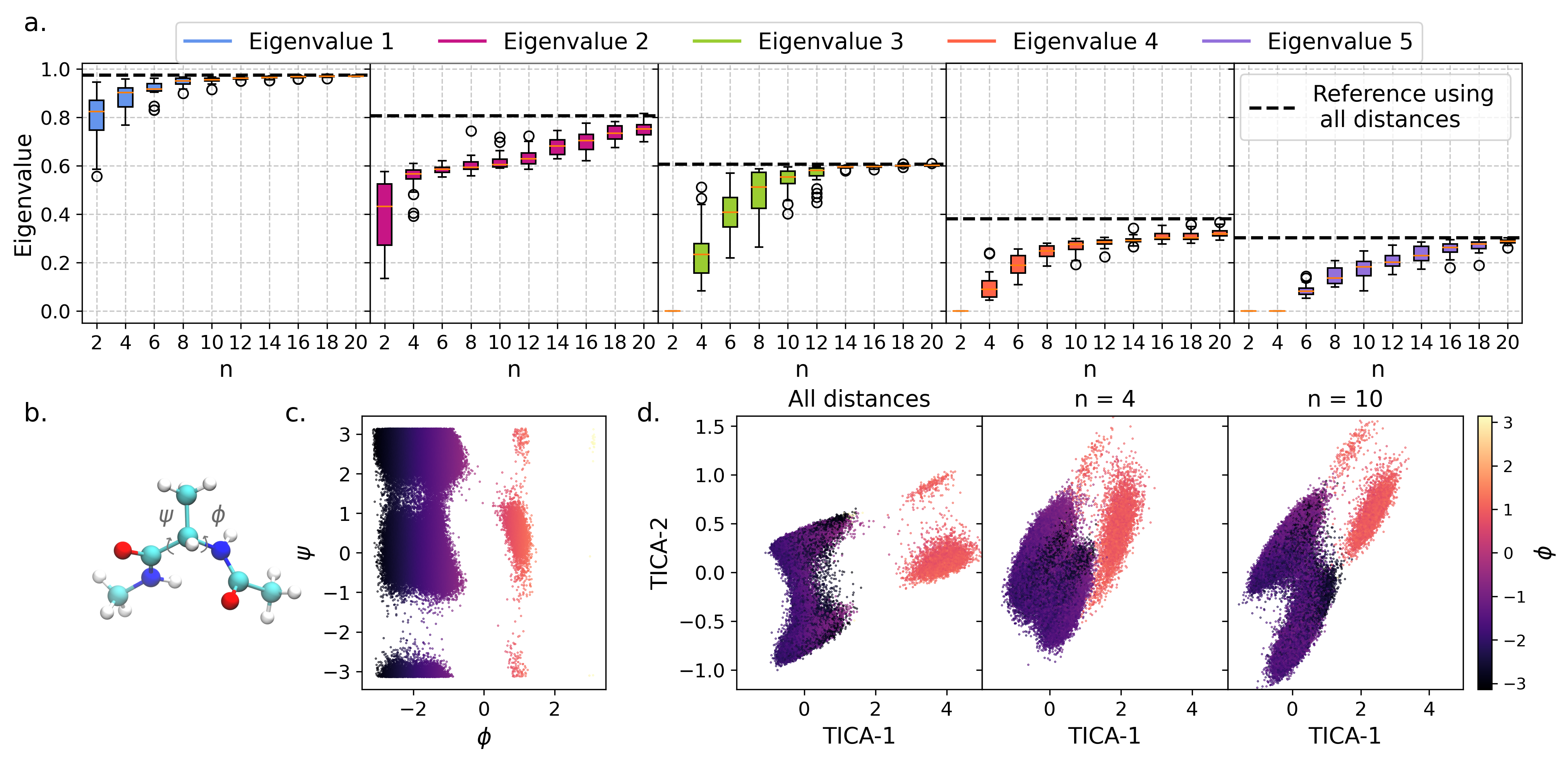}
    \caption{Alanine dipeptide results: (a) The eigenvalues of the 5 slowest TICA components (lag = 1 ps) obtained using different number of random features as input (25 trials). (b) Alanine dipeptide molecule showing the $\phi$ and $\psi$ dihedral angles. (c) Scatter plot of the dihedral angles showing the different states. (d) Examples of TICA projections obtained using different input features: all 45 distances (left), compressed dimension $n=4$ (center), compressed dimension $n=10$ (right). The scatter plots in c,d have been colored according to the $\phi$ coordinate to highlight the separation of states. }
    \label{fig:alaninedipeptide}
\end{figure}

Alanine dipeptide in aqueous solution at ambient temperature and pressure is an extremely well studied system whose dynamics is known to be captured almost entirely by the two Ramachandran angles ($\phi, \psi$). Here, we applied random projections to three independent trajectories of alanine dipeptide in TIP3P water at 300 K, each 250 ns long \cite{Timelaggedautoencoder} and available in the public repository mdshare (https://markovmodel.github.io/mdshare/).
As TICA is often used for analyzing MD simulations, we show in Figure \ref{fig:alaninedipeptide} how well the TICA components are reproduced if compressed features are used as input for these methods instead of all 45 heavy atom distances of the molecule. As randomly compressed feature sets are not unique, we generated 25 different sets of features of different dimensions taking all the distances as input for the random function generator. We then obtained TICA decompositions using these compressed features as input. In order to evaluate the quality of these decompositions, we analyzed the distributions of the first five eigenvalues over 25 trials (see Figure \ref{fig:alaninedipeptide}a). We found that a lower-dimensional compressed space was sufficient to reproduce the first TICA component, while larger dimensions were necessary for the subsequent components. With only a small set of random functions, components 2 and 3 were mixed, as were components 4 and 5, in both cases giving the smaller of the two eigenvalues (Figure \ref{fig:alaninedipeptide}a). With sufficiently many compressed features, we obtain similar TICA eigenvalues and similar projections onto the TICA components (Figure \ref{fig:alaninedipeptide}c). Even though TICA is a fairly simple linear decomposition method, it can be seen that non-linearly compressed feature sets having dimensions less than half the dimensions of the original feature set can capture both the variance and the underlying dynamics of the data set. As a nonlinear method, VAMPnet\cite{VAMPNet} improves upon many limitations of TICA and can be used to obtain both relaxation timescales and clusters from MD trajectories. 

\subsection{NTL9}
As a more demanding system, we considered NTL9, a 39-residue protein whose folding dynamics was simulated for about 1.11 ms by \citet{LindorffLarsen2011}. Recently, \citet{VAMPNet} analyzed the trajectory to obtain a Markov State Model (MSM) and timescales associated with the processes. They used 666 nearest-neighbor atom contacts defined using $c_{ij} = \exp{(-d_{ij}/d_0)}$ as input to VAMPnet, with $d_{ij}$ the pair distances and $d_0$ a characteristic length, and obtained 2-state and 5-state decompositions of the trajectory. Here, we evaluate if compressed features can produce accurate timescales and state decomposition when used as input for VAMPnet.

\begin{figure}[htbp]
    \centering
    \includegraphics[scale=0.4]{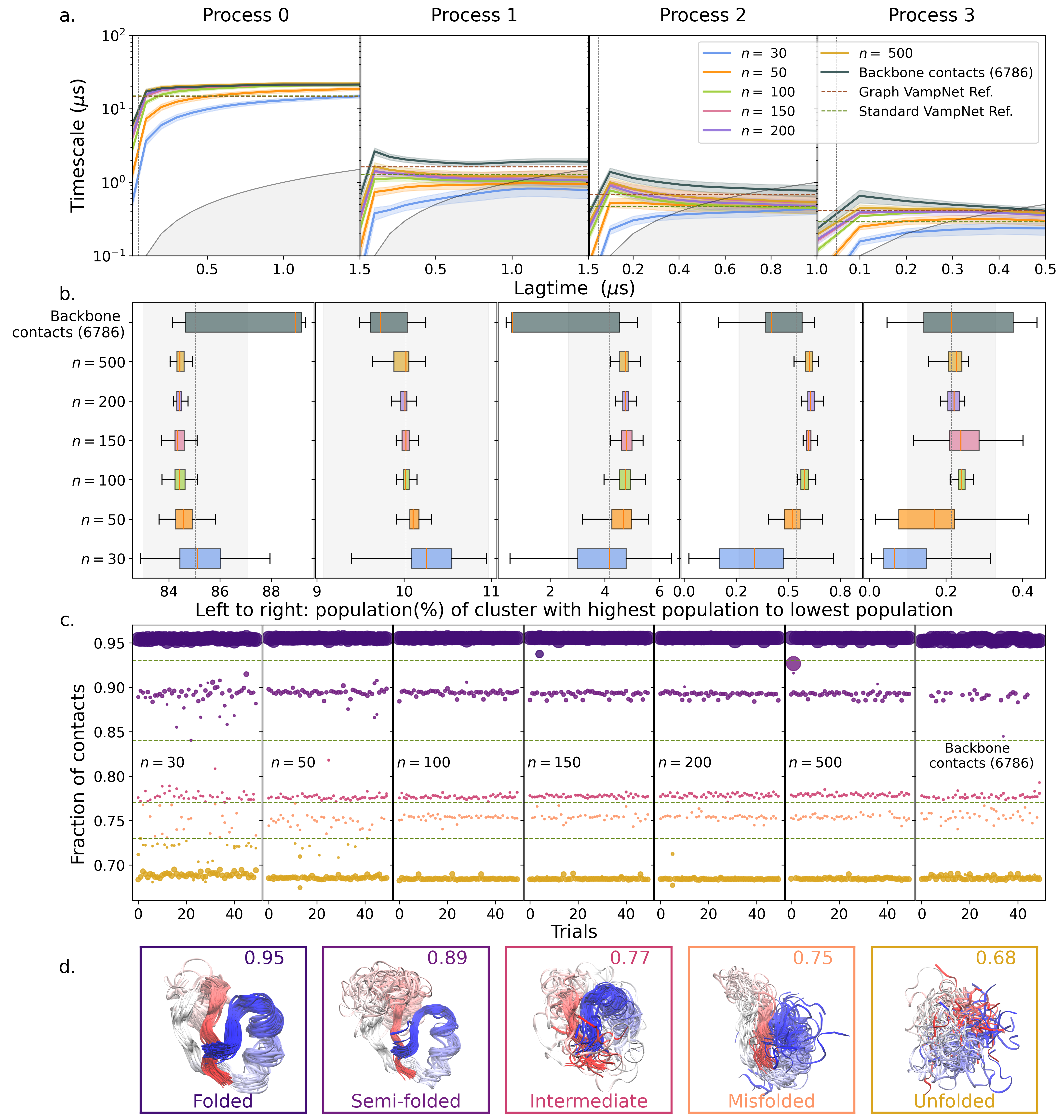}
    \caption{Random compression applied to folding of NTL9 protein:
      (a) Timescale of 4 slowest relaxation processes (left to right) extracted from 50 trials as function of lagtime, with fixed time lag of 50 ns for VAMPnet training (vertical dotted line). The dimension of the random projections, when used, are indicated by $n$ in the legend. Backbone contacts (6786) indicate that no compression is used. ``Graph VAMPnet'' are the results of \citet{Ghorbani2022-fh}, and ``Standard VAMPnet'' those of \citet{VAMPNet}. The gray area indicates timescales less than the lagtime.
      (b) Population of the five clusters obtained. Left to Right: Most populated cluster in each trial to least populated.
      (c) Mean fraction of native contacts for the five clusters obtained in the 50 trial for each method, as indicated in each subpanel. Colors correspond to the clusters in d. The size of each dot is proportional to cluster population. The four dashed lines at contact fractions of 0.93, 0.84, 0.77, and 0.73 indicate boundaries between different cluster structures.
    (d) Backbone structures representative of the clusters with different fractions of native contacts in c. The clustering was obtained in one of trials with $n = 100$ random projections. The colors of the surrounding boxes correspond to the color of the cluster in c. The value of the mean fraction of native contacts in each cluster is shown on top for reference. }
    \label{fig:ntl9}
\end{figure}

In Figure \ref{fig:ntl9}, we show the timescales and states obtained by training VAMPnets with all 6786 backbone contacts and the ones obtained using different numbers of compressed features. The compressed features were obtained from randomly chosen architectures having a randomly chosen depth between 5 and 20 layers and each layer having a random width between 2 and the input dimension. The timescales and states reported in the figure were obtained from 50 trials for each case and new random functions were generated for each trial, which then used the new compressed features as input. VAMPnets were trained using a fixed time lag of $\tau=50$ ns. However, for estimates of the relaxation time, the lagtime was varied, but with the cluster assignment fixed to that of lagtime $\tau$. The architecture of the VAMPnet lobes varied depending on the number of input features, and size of the network increased with increasing input size. 

We find that random projections of dimension $n\ge 100$ capture the slowest relaxation process of NTL9 with about the same characteristic relaxation time as obtained by using all 6786 backbone contacts (Figure \ref{fig:ntl9}a). The faster relaxation processes from the VAMPnets are somewhat slower than those from random projections, albeit with more pronounced lagtime dependence. To gain a deeper understanding, we looked at the clusters produced by the VAMPnets as representatives of the kinetic states. 
As reporters, we used the cluster population distributions and the mean fraction of native contacts for clusters. Figure \ref{fig:ntl9}b and c show the cluster population and the mean fraction of native contacts for each cluster obtained in each trial, respectively. For random projections of dimension $n\ge 100$, we find that across the respective set of 50 trials the clusters are consistent with each other, both in terms of their population (Figure \ref{fig:ntl9}b) and the extent of native structure in them (Figure \ref{fig:ntl9}c). By contrast, when using all backbone contacts in VAMPnet trials, the variation between the resulting 50 clusters is large (Figure \ref{fig:ntl9}b top and Figure \ref{fig:ntl9}c right).
We note that the five clusters correspond to the folded state, the unfolded state, and three folding intermediates (Figure \ref{fig:ntl9}c), and visually agree with the states reported by \citet{VAMPNet}, also in terms of the populations.

For NTL9, the use of high-dimensional input results in larger variation of the resulting dimensionality reduction maps in repeated trials, which may offset the finer resolution of the conformational dynamics. When all 6786 backbone contacts are used as input for VAMPnet, the populations of the clusters are distributed over a wider range of values and structures are often misclassified (Figure \ref{fig:ntl9}b,c). Furthermore, the network fails to find the third most populous semi-folded state in many trials, and multiple misfolded or unfolded clusters are found having mean fraction of native contacts between 0.73 and 0.81. By contrast, using compressed features results in a more consistent clustering as the populations of the different clusters; and the mean fraction of native contacts are consistent not only across different trials, but also across different dimensionality of compressed spaces.

Even a comparably small number of compressed features resolves the dominant processes. Although both the timescales and states are not very accurate with a compressed dimension of $n=30$, it was possible to obtain the highly populated folded and unfolded states even for this case. However, the other three states are often misclassified, as is evident from the scattered points in the mean fraction of native contacts between 0.70 and 0.80. This should also explain the significantly lower relaxation times obtained with $n=30$. However, as the dimensionality of compressed space is increased, the clustering tends to be more consistent and the relaxation times converge. As few as $n=100$ compressed features were sufficient to obtain also accurate timescales and populations (Figure \ref{fig:ntl9}a). The dimension of the compressed space ($n=100$) is significantly smaller than that of the original feature set ($6786$) or the set of 666 features used by \citet{VAMPNet}, making any analysis significantly less computationally expensive and more efficient.  
Overall, we conclude that for the NTL9 trajectory a low-dimensional compression retains the static and dynamic information encoded in the higher dimensional trajectory.

\subsection{Double-norleucin variant of villin headpiece}
Villin headpiece subdomain (HP35) is a fast-folding protein with 35 residues that has been used as a test case for many protein folding studies. One particular 300 $\mu s$ long simulation of the norleucine double mutant variant (Lys24Nle/Lys29Nle) of HP35 (PDB: 2F4K) simulated at 360 K by \citet{LindorffLarsen2011} has been studied extensively over the years\cite{Jain2014,Nagel2019,Nagel2020-vs,Nagel2023,Sormani2020-ia,damjanovic_catboss_2021,Klem2022-dp,Beauchamp2012-ab,Chang2013-oz,Chen2024-cn}. While most of these works concluded that the trajectory could be clustered into 4 states: folded, partially folded, intermediate and unfolded state, there seems to be no consensus in the literature on the exact splitting of states. For instance, \citet{Nagel2023} reported that the native basin is highly populated ($\approx 68{\%}$) while \citet{Ghorbani2022-fh} assigned only $22.93{\%}$ population to the native folded state and $71.93{\%}$ population to the unfolded state. Also, an exhaustive analysis of MSMs constructed using different input features by \citet{Nagel2023} demonstrated the necessity of feature engineering using this system. Their results indicated that selecting different types of input features, contacts or dihedrals, influenced the number of macrostates and consequently the implied timescales for different processes. The ambiguous state splitting and complicated feature selection for this system tempted us to investigate the consistency of clusters and timescales obtained using compressed features constructed with different inputs for the random function generator.

\begin{figure}[htbp]
    \centering
    \includegraphics[scale=0.38]{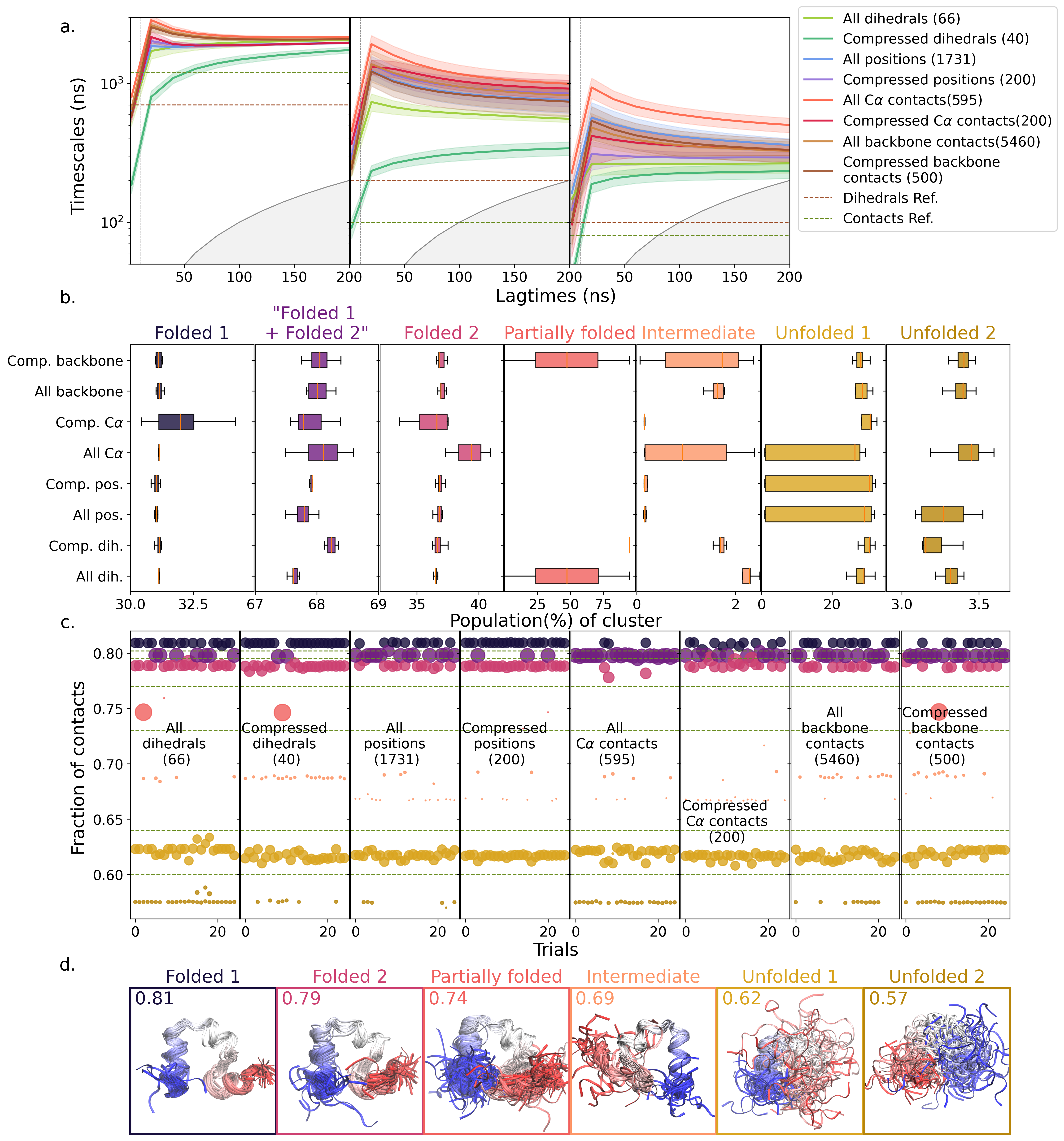}
    \caption{Random compression applied to folding of villin headpiece:
     (a) Timescale of three slowest relaxation processes (left to right) extracted from 25 trials as function of lagtime, with fixed lagtime of 20 ns for VAMPnet training (vertical dotted line). The features used, the dimension of the random projections and the method are indicated in the legend. ``Dihedrals'' and ``Contacts'' are the reference results from \citet{Nagel2019}. The gray area indicates timescales less than the lagtime.
    (b) Population of the seven clusters obtained. Clusters were grouped according to their mean number of native contacts shown in c.
    (c) Mean fraction of native contacts for the clusters obtained in the 25 trials for each method, as indicated in the figure. Colors correspond to the clusters in d.
    The size of each dot is proportional to cluster population. The six horizontal dashed lines at contact fractions of 0.802, 0.795, 0.77, 0.73, 0.64, and 0.6 indicate boundaries between different cluster structures.
    (d) Backbone structures representative of the clusters with different fractions of native contacts in c. Shown are randomly chosen representatives of each cluster across trials and methods. The colors of the surrounding boxes correspond to the color of the cluster in c. The value of the mean fraction of native contacts in each cluster is shown on top for reference. }
    \label{fig:villin}
\end{figure}

As for NTL9, we used VAMPnets to obtain the clusters and implied timescales for the double-norleucine variant of villin, which we below refer to as ``villin.'' While different types of features were used as input for VAMPnets, all networks had 3 hidden layers and 4 output neurons (4 states). We used $1.5 \times 10^5$ frames of the 300 $\mu s$ trajectory and a time lag of $\tau=20$ ns. We investigated the states obtained with 8 different sets of input features for VAMPnet: all backbone contacts (5460), all C$\alpha$ contacts (595), all dihedral angles (66), all positions (1731) and compressed features obtained using each of these 4 types of features as input to the random function generator. For the experiments with dihedral angles as input, we used the shifted dihedrals provided by \citet{Nagel2023} and for the cases with positions as input, we aligned the backbone atoms in the trajectory to those in the folded structure for all 577 atoms in villin. In Figure \ref{fig:villin}, we have summarized the VAMPnet results obtained in 25 trials for each input feature type. %while we have provided more detailed results in the SI.

In Figure \ref{fig:villin}a, we show the three slowest relaxation times obtained using different feature types. It is encouraging to note that the slowest timescales obtained in all our trials converge to similar values. Also, the timescales obtained using either complete feature sets or compressed features as input to VAMPnet are much slower than the best timescales reported in the literature\cite{Nagel2023}, and thus appear to resolve the slow dynamics well. The timescales obtained using contacts and positions are consistent with each other. By contrast, the slowest and second slowest timescales obtained using dihedral angles are somewhat faster. It is also only in the case of dihedral angles that the timescales obtained using compressed features converge to a much lower value than using the complete set of features. We conclude that positions and contacts better resolve the dynamics here than dihedrals.

To gain a structural understanding and shed light on the variations between methods, we examined the mean fraction of native contacts for each of the 4 clusters obtained across the 25 trials for different inputs. To our surprise, we observed a very different pattern for villin than what was observed for NTL9. At a first glance, we could not find any consistent 4 state split for this system using any of the input features. However, we noticed that the clusters obtained could easily be separated into 7 sets using their mean fraction of native contacts. We found that in some of our trials, the folded state was subdivided into two states (Folded1 and Folded2), each having a population of about $30{\%}$ of the population, while in other trials a single folded state with a population of about $68{\%}$ was found. This result could explain the difference in native state population observed by \citet{Nagel2023} and \citet{Ghorbani2022-fh}. Due to the very different featurization and clustering approach in these earlier studies, it seems likely that they obtained either the merged folded state or the subdivided Folded1 and Folded2 states. In addition to these folded states, we found a partially folded, an intermediate, and two unfolded states in some trials. The mean timescales obtained for different processes with different input features were therefore an average of timescales over different processes. This may explain why it was not possible to obtain a consistent 4 state splitting of states using any of the input feature sets for villin.

Nevertheless, the clusters re-grouped into 7 states have quite consistent structures (Figure \ref{fig:villin}d) and populations (Figure \ref{fig:villin}b,c). Our inability to consistently split the villin trajectory data, and the inconsistencies between published clusterings discussed above, could have multiple possible causes. The most obvious reason is that villin may have more than 4 states in the examined time regime. However, we could not get VAMPnet to converge with 7 states in this example. Another factor could be that the trajectory is not long enough to confidently determine the precise splitting of states. Despite the inconclusive splitting, Figure \ref{fig:villin}d demonstrates that running multiple trials with different feature sets can give a more fine-grained view of the mechanism of the folding process. Additionally, running multiple trials with compressed features promises a faster way to obtain different clustering solutions and gain a better idea about the number of possible states.

\section{Conclusion}
We have used neural networks for the compression of high-dimensional feature spaces of molecular dynamics trajectories. We found that random compression of the input feature spaces preserves static and dynamical information encoded in the high dimensional trajectory. We have demonstrated that when a sufficient number of random functions are used to compress the trajectory data, the implied timescales and metastable states can be reliably extracted. Having lower dimension, states and relaxation timescales tend to be more robust compared to an analysis of the full feature space.
The random features, therefore, not only reduce the need for careful feature engineering, but also offer a reliable way to reduce feature space without introducing any inherent bias. The compression of feature spaces has the potential to reduce the cost of training neural network based models for machine learning applications. They become particularly useful when high dimensionality of inputs becomes an analysis bottleneck. Interestingly, we found in our numerical trials that using $n$ independent random projections tended to produce better results than extracting the $n$ projections from one random network, as the use of independent projection networks minimizes correlations. Although we have here focused only on obtaining dimensional reduction and the construction of accurate Markov state models using VAMPnets, it is important to note that such compressed features could potentially be used as input for any machine learning model.

\begin{acknowledgement}
The authors thank the Max Planck Society for their support. The authors also thank D.E. Shaw Research for kindly providing the Villin and NTL9 folding-unfolding trajectory, Computational Molecular Biology Group at Freie University Berlin for sharing their alanine dipeptide trajectories on mdshare. The authors would also like to thank Jan F. M. Stuke, Karen Palacio Rodriguez, Sanjoy Paul and Sergio Cruz Le\'on for their comments on the manuscript.
\end{acknowledgement}

%\begin{suppinfo}

%\end{suppinfo}

\bibliography{references}

\end{document}